\documentclass[useAMS,usenatbib]{mn2e}
\usepackage{epsfig,multicol,bbm,amsmath,amssymb,euscript,array,amsfonts}
\usepackage{pstricks}
\title[UDM]{Phenomenological models for Unified Dark Matter with fast transition}

\author[M. Bruni et~al.]{Marco Bruni$^1$\thanks{marco.bruni@port.ac.uk}, Ruth Lazkoz$^2$\thanks{ruth.lazkoz@ehu.es} and
Alberto Rozas-Fern\'andez$^1$\thanks{alberto.rozas@port.ac.uk}\\
$^1$ Institute of Cosmology \& Gravitation, University of Portsmouth, Portsmouth, PO1 3FX,
UK\\
$^2$ Dpto. de F\'\i sica Te\'orica, Universidad del Pa\'is
Vasco UPV/EHU, Apdo. 644, E-48080 Bilbao, Spain}

\begin{document}

\date{\today}

\pagerange{\pageref{firstpage}--\pageref{lastpage}} \pubyear{?}

\maketitle

\label{firstpage}

\begin{abstract}
 A fast transition  between a standard matter-like era and a late $\Lambda$ cold dark matter ($\Lambda$CDM)-like epoch (or more in
general, a CDM+DE era), generated by a single Unified Dark Matter component, can provide a new interesting paradigm in
the context of general relativity, alternative to $\Lambda$CDM itself or other forms of DE or modified gravity theories
invoked to explain the observed acceleration of the Universe. UDM models with a fast transition have interesting
features, leading to measurable predictions, thus they should be clearly distinguishable from $\Lambda$CDM (and
alternatives) through observations. Here we look at different ways of prescribing phenomenological UDM models with fast
transition, then focusing on a particularly simple model. We analyse the viability of this model by studying features of
the background model and properties of the adiabatic UDM perturbations, which depend on the effective speed of sound and
the functional form of the Jeans scale. As a result, theoretical constraints on the parameters of the models are found
that allow for a behaviour compatible with observations.
\end{abstract}


\section{Introduction}

The acceleration of the expansion of the Universe is, within the homogeneous and isotropic paradigm of cosmology, a
well-accepted and observationally well-supported reality. The simplest possible framework explaining this acceleration
is provided by the concordance $\Lambda$CDM model \citep{Komatsu:2010fb}, where a cosmological constant $\Lambda$ in
Einstein equations sources the acceleration, while Cold Dark Matter (CDM) is the main component for structure formation.
 Alternatives to $\Lambda$ are various forms of Dark Energy (DE) \citep{DEbook}  or a modified gravity theory
  \citep{Nojiri:2010wj,Tsujikawa:2010zza, Clifton:2011jh}. A different approach is to consider models of Unified Dark Matter (UDM) where a single matter
component is supposed to source the acceleration and structure formation at the same time (see e.g. \cite{Kamenshchik:2001cp,Bilic:2001cg,Bento:2002ps,Carturan:2002si,Sandvik:2002jz,Scherrer:2004au,Giannakis:2005kr,Bertacca:2007ux,
Bertacca:2007cv,Bertacca:2007fc,Balbi:2007mz,Quercellini:2007ht,Pietrobon:2008js,Bertacca:2008uf,Bilic:2008yr,Camera:2009uz,Li:2009mf,
Piattella:2009kt,Gao:2009me,Camera:2010wm,Lim:2010yk}).

In a previous paper \citep{Piattella:2009kt} the concept of UDM models with fast transition was introduced (see
\cite{Bassett:2002qu} for DE models with a sharp transition in their equation of state). In essence, these models are
based on the idea that the Universe may have undergone a transition between a standard matter-like era, well described
by an Einstein-de~Sitter (EdS) model, and a $\Lambda$CDM-like epoch. If the transition is slow the differences with
$\Lambda$CDM are negligible while if the transition is fast, these models show interesting features\footnote{In a
$\Lambda$CDM model of course there is an early matter era when $\Lambda$ is negligible, with a smooth slow transition to
the epoch when $\Lambda$ plays a role. We are instead interested in models where there is a longer matter era that
almost suddenly ends into a $\Lambda$CDM-like late behaviour; in other words, models where the UDM component evolves
for long time in a CDM-like fashion, and suddenly a $\Lambda$-like term is switched on in the dynamics.}
\citep{Piattella:2009kt,Bertacca:2010mt}. The transition can be quantified as fast by looking both at parameters that
govern the evolution of the background as well as at quantities that dictate the dynamics of perturbations. This
analysis has been carried out in \cite{Piattella:2009kt} for a specific barotropic model and will be generalised in this
paper to a new class of models, but in essence the transition needs to be fast because: {\it a)} we are especially
interested in background models that, at least in principle, can be clearly distinguished from a standard  $\Lambda$CDM;
{\it b)} as shown in \cite{Piattella:2009kt}, otherwise the evolution of perturbations is such that observational
constraints are violated, in particular causing a strong deviation from the Integrated Sachs-Wolfe (ISW) effect
occurring in $\Lambda$CDM models.

The UDM models with fast-transitions may be an interesting alternative to other explanations of the observed
acceleration of the Universe, providing a good fit to standard observables such as the CMB and the matter distribution
\citep{Piattella:2009kt} that require consideration of the  perturbations, while at the same time showing interesting
new features, leading to measurable predictions. So at least in principle, UDM models with a fast transition are able to avoid the fate of some UDM models such as the generalised Chaplygin gas, which need to become indistinguishable from $\Lambda$CDM in order to survive observational tests, which spells their end \citep{Sandvik:2002jz}, cf.\
\citep{Gorini:2007ta,Piattella:2009da}. On the other hand, UDM models based on scalar fields, such as the one introduced in \cite{Bertacca:2008uf}, can be compatible with observations. See also \cite{Bertacca:2010ct} for a recent review on UDM models.

  In contrast with more standard CDM+DE models, where the CDM component is perturbed and leads to structure formation
while the DE component takes care of the acceleration of the background, with small or negligible effects on
perturbations, the single UDM component must accelerate the Universe and provide acceptable perturbations. In
particular, while CDM density perturbations evolve in a scale-independent fashion, this is not the case for UDM. In
view of testing models against observations, e.g.\ with Markov Chain Monte Carlo methods and likelihood analysis, these
differences may become computationally  expensive. In particular, modifying CAMB \citep{Lewis:1999bs} to treat fast
transition UDM models implies to switch off CDM while introducing a rapidly varying single inhomogeneous component with
scale dependent evolution. It is then a non trivial task to obtain a working code that it is efficient enough for likelihood analysis \citep{Piattella:2009kt}, given that the running time of a code like CAMB \citep{Lewis:1999bs}, when dealing with a non-standard model like a fast transition UDM, increases enormously when the accuracy is increased in order to retain convergence of the results.

      In view of this, and lacking a fundamental model, it is therefore essential to consider simple phenomenological models
of the fast-transition paradigm for which as much theoretical progress as possible can
be made from analytical calculations. This then can be used to increase the efficiency of numerical codes such as CAMB
\citep{Lewis:1999bs} and CLASS \citep{Lesgourgues:2011re} in dealing with these models.

 Our first goal here is therefore to look at simple phenomenological recipes,
such that the most important variables required for the numerical problem can be expressed analytically. It then
turns out that, unlike the fast transition UDM model introduced in  \cite{Piattella:2009kt}, where the prescription for
the fast transition is introduced in the equation of state, the best recipe to proceed analytically is to prescribe
the evolution of the energy density of UDM. We then introduce a specific simple new model based on this recipe and analyse its properties in some detail, establishing what its range of validity should be if compared if observations. We find similar results to those
in  \cite{Piattella:2009kt} but, given the very different starting points, this is in itself a non trivial outcome.

The rest of the paper can be outlined as follows. In section \ref{sec:udmmodels} we introduce the basic equations
describing the background and the perturbative evolution for a general UDM model. In section \ref{sec:3p} we explore
three possible prescriptions for the dynamics of the UDM component. In section \ref{sec:tghmodel} we introduce a new
UDM model with fast transition and study its background evolution, comparing it to a $\Lambda$CDM. In section
\ref{sec:perts} we analyse the properties of perturbations in this model, focusing on the evolution of the  effective
speed of sound and that of the Jeans length during the transition. The conclusions are drawn in section
\ref{sec:concl}.


\section{Generalities of UDM models}\label{sec:udmmodels}



\subsection{The background}

 We assume  a spatially flat Friedmann-Lema\^{\i}tre-Robertson-Walker (FLRW) cosmology, although many of the
considerations in the following sections would apply in a more general context. The metric then is $ds^2 = -dt^2
+a^2(t)\delta_{ij}dx^{i}dx^{j}$, where $t$ is the cosmic time, $a(t)$ is the scale factor and $\delta_{ij}$
is the Kronecker
delta. The total stress-energy tensor is that of a  perfect fluid with energy density  $\rho$
and pressure $p$,  with $u^{\mu}$ its four-velocity: $T_{\mu\nu} = \left(\rho + p\right)u_{\mu}u_{\nu} + pg_{\mu\nu}$.
Starting from these assumptions, and choosing units such that $8\pi G = c = 1$ and signature $\{-,+,+,+\}$, Einstein
equations imply the Friedmann and Raychaudhuri equations:
\begin{eqnarray}\label{Fried}
 H^2 & =& \left(\frac{\dot{a}}{a}\right)^2 = \frac{\rho}{3}\;,\\
\label{Raycha}
 \frac{\ddot{a}}{a} &=& -\frac{1}{6}\left(\rho + 3p\right)\;,
\end{eqnarray}
where $H=\dot{a}/a$ is the Hubble expansion scalar and the dot denotes derivative with respect to the cosmic time.
Assuming that the energy density of the radiation is negligible at the times of interest, and disregarding also the
small baryonic component, $\rho$ and $p$ represent the energy density and the  pressure of the UDM component.

Independently from Einstein equations, projecting the conservation equations $T^{\mu\nu}{}_{;\nu}=0$ along $u^{\mu}$
one obtains the energy conservation equation
\begin{equation}\label{enconseq}
 \dot{\rho} = -3H\left(\rho + p\right) = -3H\rho\left(1 + w\right)\;,
\end{equation}
where $w = p/\rho$ represents the equation of state (hereafter
EoS) that is needed to close the system and is the quantity that
characterises the background of our UDM model.

When needed, we shall introduce different components, each with energy density $\rho_{i}$. From this, we can define the
dimensionless function

\begin{equation}\label{omegas}
\Omega_{\rm i}(a)=\frac{\rho_{\rm i}(a)}{\rho_{\rm cr}(a)}\;,
\end{equation}
 where $\rho_{\rm cr}=3H^{2}$ is the critical density; values today will be denoted by the parameter $\Omega_{\rm
i,\rm 0}$.


\subsection{Perturbations}


Assuming a perfect fluid,  perturbations of the FLRW metric in the longitudinal gauge read
\begin{equation}
\label{pertmetric}
ds^2 = -a^2(\eta)\left[\left(1 + 2\Phi\right)d\eta^2 - \left(1 - 2\Phi\right)\delta_{ij}dx^idx^j\right]\;,
\end{equation}
where $\Phi$ represents the analogous of the Newtonian gravitational potential and we now are  using conformal time
$\eta$.

Defining
\begin{equation}
u = \frac{2\Phi}{\sqrt{\rho + p}}\;
\end{equation}
and linearising the 0-0 and 0-i components of Einstein equations,  one obtains the following second order differential
equation  for the Fourier component of  $u$  \citep{Mukhanov:1990me,Giannakis:2005kr,Bertacca:2007cv,Piattella:2009kt}:
\begin{equation}\label{equ}
\frac{d^{2}u}{d\ \eta^{2}} + k^{2}c_{\rm s}^{2}u - \frac{1}{\theta} \frac{d^{2}\theta}{d\ \theta^{2}} u = 0\;,
\end{equation}
where
\begin{equation}\label{theta}
 \theta = \sqrt{\frac{\rho}{3(\rho + p)}}(1 + z)\;,
\end{equation}
with  $z$  the redshift, $1 + z = a^{-1}$.
The quantity
 $c_{\rm s}^2$ in (\ref{equ})  is the effective speed of sound and characterises the perturbative dynamics of our UDM
model, being also crucially involved in the growth of the overdensities $\delta\rho$. Assuming adiabatic perturbations
this is the same as the adiabatic speed of sound:
\begin{equation}
\label{c1} c_{\rm s}^{2}=c_{\rm ad}^2=\frac{dp}{d\rho}=\frac{ \frac{dp}{d \eta} }{\frac{d\rho}{d\eta}}\;.
\end{equation}

Starting from Eq. (\ref{equ}), let us define the squared Jeans wave number:
\begin{equation}
 k^{2}_{\rm J} = \left|\frac{1}{c_{\rm s}^{2}\theta} \frac{d^{2}\theta}{d\theta^{2}\,}  \right|\;;
\end{equation}
its reciprocal  defines the squared Jeans length: $\lambda^2_{\rm J} = a^{2}/k^2_{\rm J}$.

An important aspect of UDM models is the possible manifestation of
an effective sound speed significantly different from zero at late
times: this generally corresponds to the appearance of a Jeans
length (or sound horizon) below which the dark fluid does not
cluster (e.g. see
\cite{Hu:1998kj,Pietrobon:2008js,Piattella:2009kt}). This causes a
strong evolution in time of the gravitational potential, which at small scales
starts to oscillate and decay, with effects  on  structure formation. In general, UDM models may also exhibit  a strong
 ISW effect \citep{Bertacca:2007cv}.

Thus, the squared Jeans wave number plays a crucial role in determining the viability of a UDM model, because of its
effect on perturbations, which is then revealed in observables such as the CMB and matter power spectrum
\citep{Pietrobon:2008js,Piattella:2009kt}.  As  discussed in \cite{Piattella:2009kt} there are two different regimes of
evolution, respectively for scales much smaller and much larger than the Jeans length. In practice, any viable UDM
model should satisfy the condition  $k_{\rm J}^2 \gg k^2$ for all the scales of cosmological interest, in turn giving
an evolution  for the gravitational potential $\Phi$ in Fourier's space  of the following type (we are dealing with the
gravitational potential after recombination so there is no more speed of sound due to radiation):
\begin{equation}\label{kjggksol}
 \Phi(\eta,k) \simeq A_{\rm k}\left[1 - \frac{H(\eta)}{a(\eta)}\int a(\hat{\eta})^2 d\hat{\eta}\right]\;.
\end{equation}
The integration constant  $A_{\rm k} = \Phi\left(0,k\right)T_{\rm m}\left(k\right)$  is fixed during inflation by the
primordial potential  $\Phi\left(0,k\right)$ at  large scales;  $T_{\rm m}\left(k\right)$ is the matter transfer
function, describing the evolution of perturbations through the epochs of horizon crossing and radiation-matter
transition, see e.g.\ \cite{Dodelson:2003ft}.

 The explicit form of the Jeans wave number is
 \begin{multline}\label{kJ2analytic}
 k_{\rm J}^{2}  = \frac{3}{2}\rho a^{2} \frac{(1 + w)}{c_{\rm s}^2}\left|\frac{1}{2}(c_{\rm s}^2 - w) -
\rho\frac{dc_{\rm s}^2}{d\rho} + \right.\\
\left.\frac{3(c_{\rm s}^2 - w)^2 - 2(c_{\rm s}^2 - w)}{6(1 + w)} + \frac{1}{3}\right|\;.
\end{multline}
It is therefore  clear from this expression that, if we want an
analytic expression for $k_{\rm J}^{2}$ in order to obtain some
insight on the behaviour of perturbations in a given UDM model, we
need to be able to obtain analytic expressions for $\rho$, $p$,
$w$ and $c_{\rm s}^{2}$. Unfortunately, it is not possible to find
such expressions  as functions of $\eta$ (or $t$), simply because
this requires the knowledge of an analytic expression for the
scale factor as function of time, i.e.\ to solve  the Friedmann
equation (\ref{Fried}), which in general is only possible for very
special cases, as is well known.


\section{Three possible prescriptions}\label{sec:3p}

A way out of this problem is to disentangle the evolution of the quantities of interest  ($\rho$, $p$, $w$ and
$c_{\rm s}^{2}$) from Einstein equations, noticing that we can obtain these quantities as functions of the scale factor
$a$
if we use only the conservation equation (\ref{enconseq}). This has also the advantage that the expressions so obtained
will be the same in any theory of gravity that satisfies the conservation equations.  Eq.\ (\ref{enconseq}) becomes,
using $a$ as time variable:

\begin{equation}
\label{encon1}
\rho' =-\frac{3}{a}(\rho +p)\;,
\end{equation}
where a prime indicates derivative with respect to $a$.

We now briefly look at three different possible ways to prescribe the dynamics of the UDM component and to derive
analytic expressions for the needed variables.

\subsection{Starting from  $w(a)$}


Suppose that $p/\rho=w$ is pre-assigned as a function of the scale factor: $w=w(a)$. For instance, $w(a)=w_0+w_a(1-a)$
is a typical phenomenological assumption well-motivated when setting observational constraints on dark energy models
\citep{Chevallier:2000qy,Linder:2002et}. In principle this is a convenient practical prescription to model UDM, because
we have a good idea about what type of $w(a)$ we should have. Then the adiabatic speed of sound
\begin{equation}
\label{c2}
c_{\rm s}^{2}=\frac{dp}{d\rho}=\frac{\frac{dp}{da}}{\frac{d\rho}{da}}=\frac{p'}{\rho'}
\end{equation}
can be computed from the conservation equation (\ref{encon1}).
Indeed, from $p=w\rho$ we can compute $p'$,  then substituting the latter and $\rho'$ from (\ref{encon1}) to obtain
\begin{equation}
\label{c3}
c_{\rm s}^{2}=w -\frac{aw'}{3(1+w)}\;.
\end{equation}
This prescription however doesn't lead to analytic expressions for $\rho(a)$ and $p(a)$ in general, unless
$\int\frac{1+w(a)}{a}da$ is integrable.


\subsection{Starting from  $p(a)$}


Prescribing the pressure as a function of the scale factor, $p=p(a)$, can be useful, e.g.\ if one is dealing with a
scalar field, in which case this is equivalent to prescribing a Lagrangian, see e.g.\
\cite{Bertacca:2007ux,Quercellini:2007ht,Bertacca:2010ct} and Refs.\ therein.  Again, we have a good idea about the
functional form that $p(a)$ should
have, so that this also seems a good starting point. Let's rewrite the energy conservation (\ref{encon1}) as
\begin{equation}
\label{encon2}
\rho' +\frac{3}{a}\rho =-\frac{3}{a}p(a)\;.
\end{equation}
The homogeneous solution is $\rho_{m}\propto a^{-3}$, i.e.\ standard matter (dust), and for a given $p(a)$ an analytic
expression for $\rho(a)$ can be found if $ E= 3 \int a^{2} p(a) da=\int p dV$ is integrable, giving\footnote{The
expansion is adiabatic, and $\rho_{m}\propto V^{-1}$, thus we can interpret $E$ as the energy of the system in the
volume $V$.} $\rho=E/V$. With this prescription $c_{\rm s}^{2}$ is immediately found, given $p(a)$ and (\ref{encon2}),
but
an analytic expression for  $w(a)$ can only be found  if that for $\rho(a)$ is found.


\subsection{Starting from $\rho(a)$}


It is perhaps less obvious what the functional form for $\rho(a)$ should be, but some guess can be made in view of
constructing UDM models with fast transition. In this case we want to recover a CDM-like behaviour, i.e.\ an EdS model,
at early times (before the transition), i.e.\ $\rho\simeq \rho_{\rm m}=\rho_{\rm M,\rm 0}a^{-3}$, and a CDM+DE behaviour
after the
transition. If we want to recover the simplest case of a $\Lambda$CDM at late times, we would have $\rho\simeq
\rho_{\Lambda}+\rho_{\rm M,\rm 0}a^{-3}$.

Assuming that  $\rho(a)$ is given, we can compute $\rho^{\prime}(a)$ and then use (\ref{encon2}) to obtain $p(a)$ and
then $p'(a)$. In this case therefore {\it any} choice of $\rho(a)$ guaranties analytic expressions for the EoS $w(a)$
and the adiabatic speed of sound $c_{\rm s}^{2}(a)$.

To summarise, given a function (at least of class $C^{3}$) $\rho=\rho(a)$ for the energy density,  we have the
following expressions for the quantities that enter into the Jeans wave number (\ref{kJ2analytic}):
\begin{eqnarray}\label{varie1}
w & = & - \frac{a}{3}\, \frac{\rho'}{\rho}-1\;, \\
c_{\rm s}^{2} & = & - \frac{a}{3}\,\frac{\rho''}{\rho'} -\frac{4}{3}\;,
\label{varie2} \\
\frac{dc_{\rm s}^{2}}{d\rho}  & = & - \frac{1}{3\rho'^2}\, \left[
a \rho'''+\rho''-a \frac{\rho''^2}{\rho'}\right]\;. \label{varie3}
\end{eqnarray}
Substituting in Eq.\ (\ref{kJ2analytic}) from (\ref{varie1})-(\ref{varie3}) we can obtain an analytic expression for
the function $  k_{\rm J}^{2}(a)$. Armed with this, we can obtain some insight about the behaviour of adiabatic
perturbations in a model with a specified energy density $\rho=\rho(a)$.


\section{Phenomenological UDM models with fast transition}\label{sec:tghmodel}

\subsection{An overidealised model}

Before looking at a possible model for a UDM with fast transition, we now introduce an overidealised model, using a
Heaviside function to describe an instantaneous transition. This model can't serve our purposes, because it is clear
from the expressions above that we need the function $\rho(a)$ to be  at least of class $C^{3}$,
but it is useful to get an idea of what we want to obtain. We assume  that the Universe is well described by an EdS
model before the transition, while for generality we  describe the post-transition era with an ``affine" model
\citep{Ananda:2005xp,Ananda:2006gf,Balbi:2007mz,Quercellini:2007ht,Pietrobon:2008js}:
 \begin{equation}
\label{affine}
\rho=\left\{
\begin{array}{ll}
  \rho_{\rm t}\left(\frac{a_{\rm t}}{a}\right)^{3}    &   a<a_{\rm t} \\
\rho_{\Lambda} + (  \rho_{\rm t} - \rho_{\Lambda} ) \left(\frac{a_{\rm t}}{a}\right)^{3(1+\alpha)}     &   a>a_{\rm t}
\end{array}\right.
\end{equation}
Of course, any other post-transition model could be chosen.
Here $\rho_{\rm t}$ is the  energy scale at the transition and
$\rho_{\Lambda}$ is the effective cosmological constant: $1+\alpha>0$ and the energy density at late times tends to
$\rho_\Lambda$, so that the late time evolution in these models is {\it a-la} de~Sitter, even if there is no
cosmological
constant in the Friedmann and Raychaudhuri equations, see
\cite{Ananda:2005xp,Ananda:2006gf,Balbi:2007mz,Pietrobon:2008js}. On the other hand, $a_{\rm t}$ is not an independent
parameter: using the Friedmann equation (\ref{Fried}) and the
second of Eq.\ (\ref{affine}), and neglecting radiation, we get
\begin{equation}\label{at}
a_{\rm t}=\left[\frac{1-\Omega_{\Lambda,0}}{\Omega_{\Lambda,0}\left(\frac{\Omega_{\rm
t,0}}{\Omega_{\Lambda,0}}-1\right)} \right]
^{\frac{1}{3(1+\alpha)}}
\end{equation} where $\Omega_{\rm t,\rm 0}/\Omega_{\Lambda,0}=\rho_{\rm t}/\rho_{\Lambda}$; this gives a redshift for
the
transition $z_{\rm t}=a_{\rm t}^{-1}-1$.
  We can then interpret our UDM model after the transition as made up of the
effective cosmological constant $\rho_{\Lambda}$ and an evolving
part, with energy density at the transition $\rho_{\rm m}\equiv\rho_{t} -
\rho_{\Lambda}$,\footnote{In this two component interpretation the two energy densities
satisfy their own conservation equations and therefore there is no
coupling between them.} that decreases after the transition.

In the affine model the energy density of UDM is given by the second of Eq.\  (\ref{affine}). If one assumes this model
all the way deep into the radiation era, the evolving part must have
$w_{\rm m}\rightarrow0$ and $c_{\rm s}^{2}\rightarrow0$ in the past in
order to recover standard matter domination at early times,  thus the evolving part can be interpreted as the "dark
matter" component today. Indeed, the value of the parameter $\alpha$ is
extremely constrained if one assumes an affine model with
adiabatic perturbations all the way back to recombination and
beyond, $\alpha\approx 10^{-7}$ \citep{Pietrobon:2008js} (cf.\ also
\cite{Muller:2004yb}), making this model indistinguishable from
$\Lambda$CDM.\footnote{The $\Lambda$CDM can be obtained as a
sub-case of the UDM affine model for $\alpha=0$ \citep{Ananda:2005xp,Ananda:2006gf,Balbi:2007mz}; the constraint is
weaker if the perturbations are not adiabatic,  see
\cite{Pietrobon:2008js}.} However,
such a strong bound mainly comes from the matter power spectrum at
small scales, $k\gtrsim k_{\rm J}$, where the growth of
perturbations is affected by a non-vanishing $k_{\rm J}$. Since
this is an integrated effect, it is reasonable to expect that, if
the affine-like evolution only starts below a transition red-shift $z_{\rm t}$, the
bound will be much weaker.

\begin{figure}
\begin{center}
\includegraphics[width=0.99\hsize]{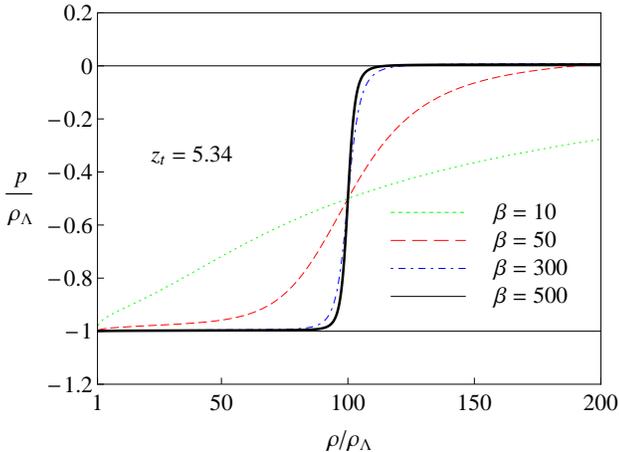}
\caption{Illustrative parametric plot of $p/\rho_\Lambda$ as a function of $\rho/\rho_\Lambda$ with
$\rho_{\rm t}/\rho_\Lambda = 100$ ($z_{\rm t}\thicksim5.34$) for the model specified by Eq.~(\ref{Hc}). The $\Lambda$CDM
model is represented
here by the solid horizontal line $p/\rho_\Lambda=-1$, while the line $p=0$ represents an EdS model, i.e. pure CDM. All
models asymptotically evolve towards the effective cosmological constant $\rho_\Lambda$.}
\label{pvsrhoFig1}
\end{center}
\end{figure}

It is useful  to explicitly incorporate a Heaviside function $H(a-a_{\rm t})$ in Eq.~(\ref{affine}), so that:
\begin{equation}
\label{step}
\rho=  \rho_{\rm t}\left(\frac{a_{\rm t}}{a}\right)^{3} +\left[ \rho_{\Lambda} +(\rho_{\rm t}-\rho_{\Lambda})
\left(\frac{a_{\rm t}}{a}\right)^{3(1+\alpha)} -\rho_{\rm t}\left(\frac{a_{\rm t}}{a}\right)^{3} \right]
H(a-a_{\rm t})\;.
\end{equation}
For $\alpha=0$ this reduces to
\begin{equation}
\label{step1}
\rho=  \rho_{\rm t}\left(\frac{a_{\rm t}}{a}\right)^{3} +  \rho_{\Lambda} \left[ 1- \left(\frac{a_{\rm t}}{a}\right)^{3}
\right]
H(a-a_{\rm t})\;,
\end{equation}
 representing  a sudden transition to $\Lambda$CDM. In the following, we shall restrict our attention to this sub-class
of models.

 It is now clear that, simply replacing $H(a-a_{\rm t})$ with a smoother transition function  $H_{\rm t}(a-a_{\rm t})$,
we can
obtain simple UDM models with a fast transition.

\subsection{A simple model for the background}

In this paper, among the many known continuous approximations to the Heaviside function \citep{Bracewell}, we shall
consider the only one that we found compatible with having $c_{\rm s}^{2}>0$:
\begin{equation}
\label{Hc}
H_{\rm t}(a-a_{\rm t})=\frac{1}{2} + \frac{1}{\pi}\arctan(\beta(a-a_{\rm t})),
\end{equation} where the parameter $\beta$ represents the rapidity of the transition. In addition to $a_{\rm t}$ and
$\beta$,
 there is a third parameter in the model, which is $\rho_{\Lambda}$, or, equivalently, the corresponding density
parameter $\Omega_{\Lambda,0}$. From inserting Eq.~(\ref{Hc}) in  Eq.~(\ref{step1}) we see that asymptotically in time,
in the limit $a\rightarrow\infty$, $\rho\rightarrow\rho_{\Lambda}$, which implies  $p\rightarrow-\rho_{\Lambda}$. As
already mentioned above, $\rho_{\Lambda}$ plays the role of an effective cosmological constant, i.e. it is an attractor
for Eq.~(\ref{encon1}). The Universe necessarily evolves toward  an asymptotic de~Sitter phase, i.e. a sort of cosmic
no-hair theorem holds (see  \cite{Bruni:2001pc,Bruni:1994cv} and refs.\ therein and
\cite{Ananda:2005xp,Ananda:2006gf,Balbi:2007mz}).
In Fig.\ \ref{pvsrhoFig1} we show a parametric plot of $p$ vs $\rho$,  normalised to $\rho_{\Lambda}$, where we have
assumed, as an example, that the transition takes place at $z_{\rm t}=5.34$,  for a representative choice of values of
$\beta$. It can be seen that all models gradually approach the effective cosmological constant $\rho_\Lambda$.

\begin{figure}
\begin{center}
\includegraphics[width=1.13\columnwidth]{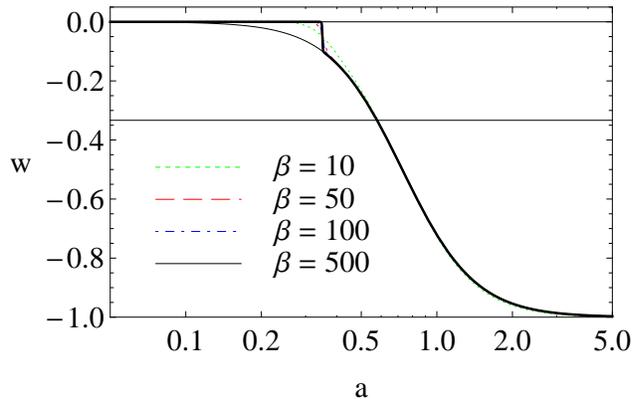}
\includegraphics[width=1.13\columnwidth]{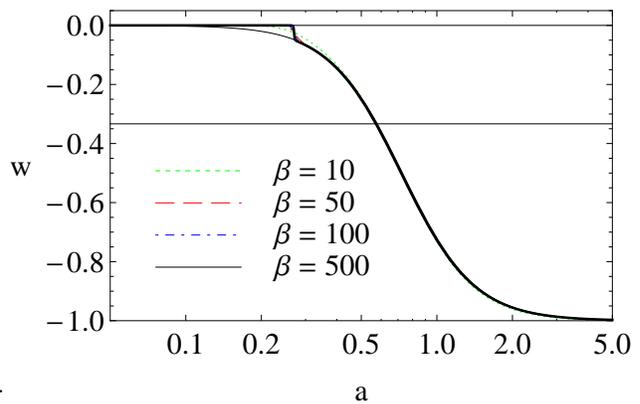}
\caption{Evolution of the UDM parameter $w = p/\rho$ as a function of the scale factor $a$ for $\rho_{\rm
t}/\rho_\Lambda = 10$ ($a_{\rm t}\thicksim0.35$, $z_{\rm t}\thicksim1.85$) (top panel) and $\rho_{\rm t}/\rho_\Lambda =
20$
($a_{\rm t}\thicksim0.27$, $z_{\rm t}\thicksim2.65$) (bottom panel). For reference we also plot: the $w = 0$ line
representing a
flat
pure CDM model (an EdS universe); the $w=-1/3$ line representing the boundary between the decelerated and the
accelerated phases and the thin black curved line representing the evolution of the total $w$ for the $\Lambda$CDM
model with $\Omega_{\Lambda,0} = 0.72$. The higher $\rho_{\rm t}/\rho_\Lambda $, the earlier the UDM $w$ transits to
that of a $\Lambda$CDM. In the future, for $a>1$, all models evolve to a de~Sitter phase with $w=-1$.}
\label{wvszFig2}
\end{center}
\end{figure}

In order to make our UDM model close to  $\Lambda$CDM  at late times, the transition must occur at relatively high
redshifts,  such that $\rho_{\rm t}$ is quite larger than $\rho_\Lambda$, which corresponds to a minimum value  of the
redshift $z_{\rm t}$. Otherwise, it could be difficult to have a good fit of supernovae and ISW effect data
\citep{Piattella:2009kt}. For instance, we need $z_{\rm t} \gtrsim 2.65$ if we want to have  $\rho_{\rm t} \gtrsim
20\rho_\Lambda$.
In Fig.\ \ref{wvszFig2} the evolution of the EoS $w$ is depicted as a function of the scale factor for different values
of $\rho_{\rm t}/\rho_\Lambda$ and $\beta$.
As shown, models with a smaller rapidity $\beta$ have a background evolution  more similar to that of the $\Lambda$CDM
model at all times.
On the other hand, a larger $\beta$ implies a sharper transition between the CDM-like phase and the $\Lambda$CDM phase.
Likewise, it clearly illustrates that the  transition has to take place far enough in the past, i.e. $\rho_{\rm t}$ is
larger than $\rho_\Lambda$, in order for the late time evolution of $w$ to be  close to that of the  $\Lambda$CDM model.
\begin{figure}
\begin{center}
\includegraphics[width=1.24\columnwidth]{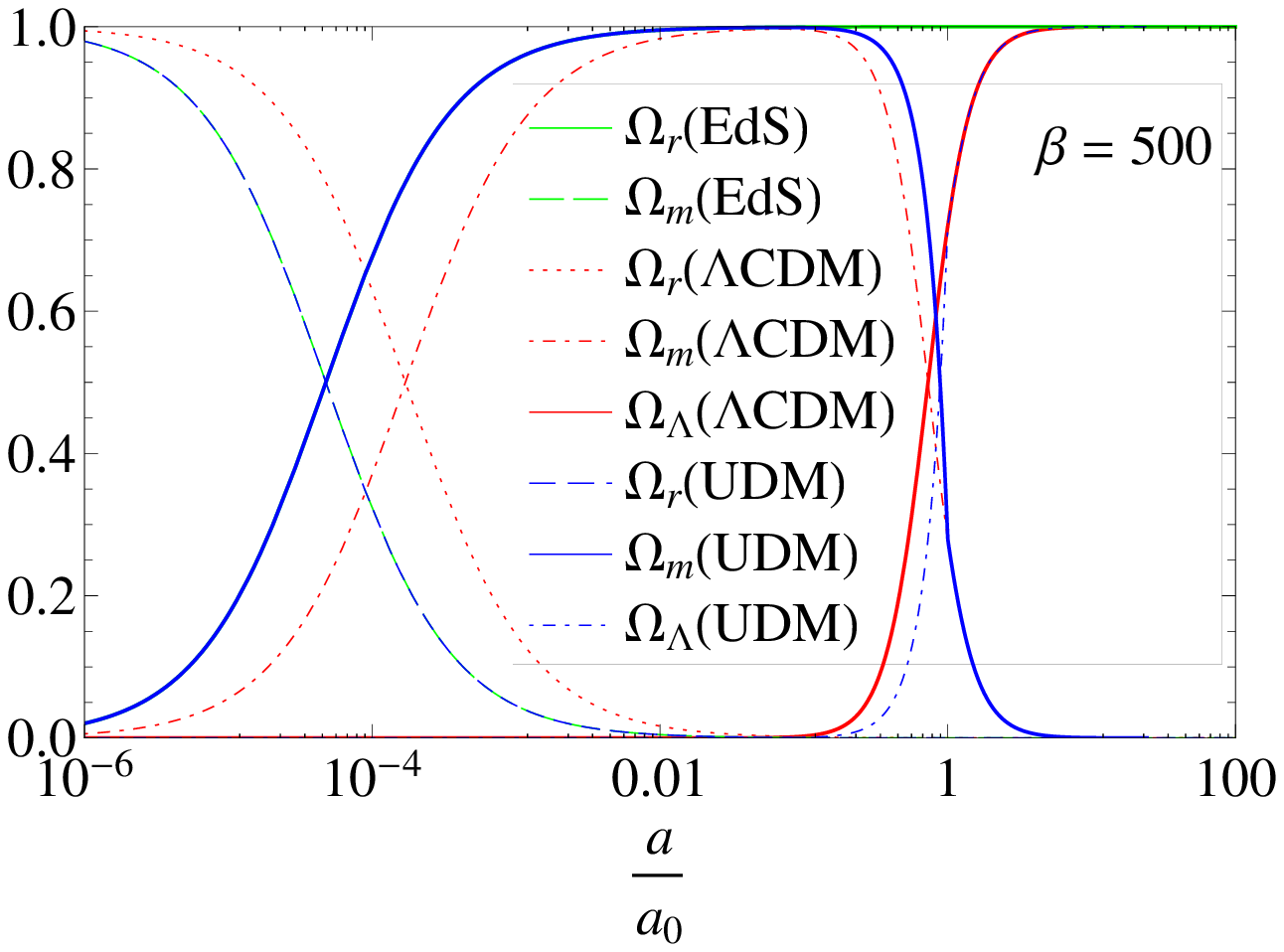}
\includegraphics[width=1.24\columnwidth]{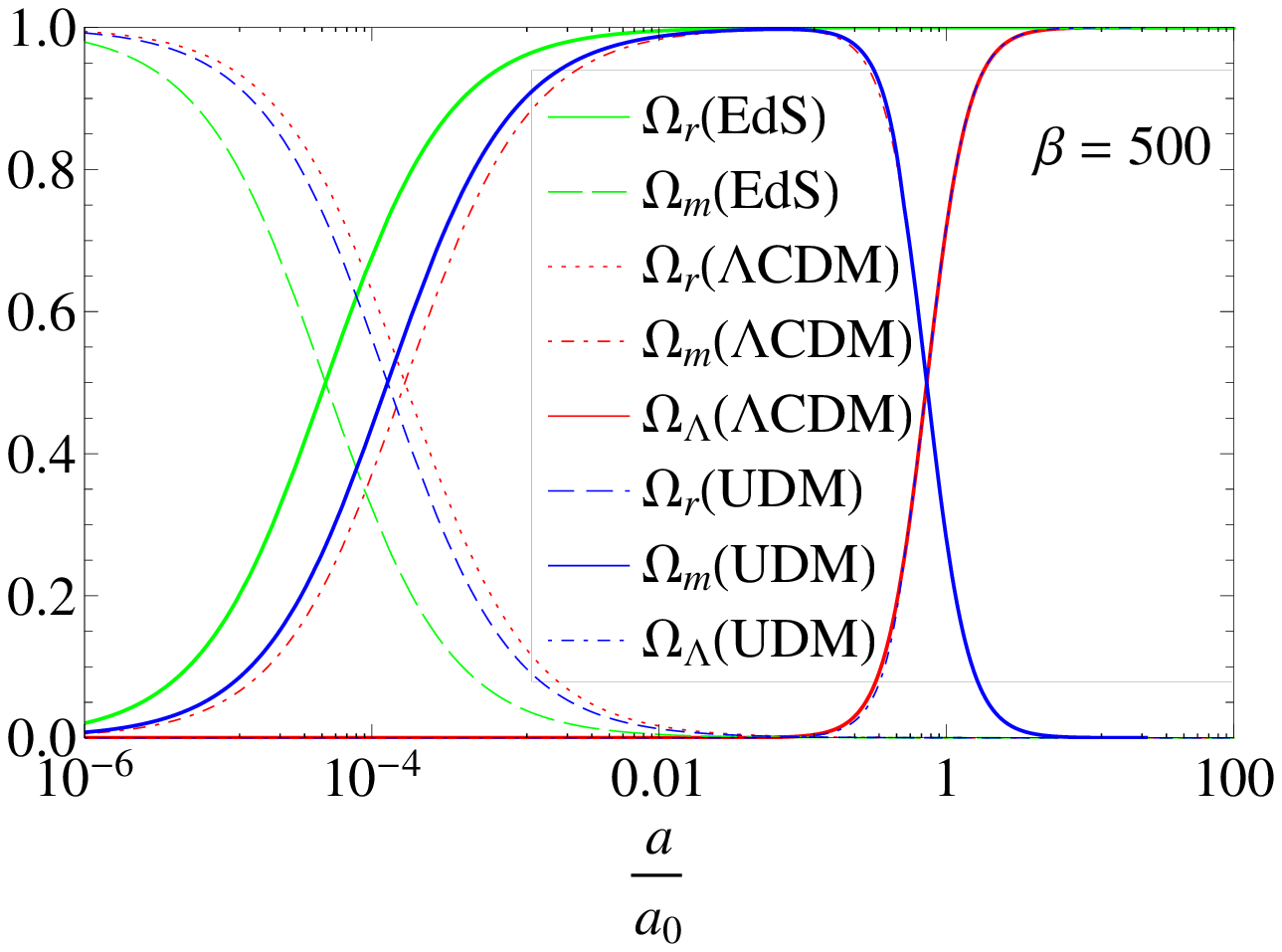}
\includegraphics[width=1.24\columnwidth]{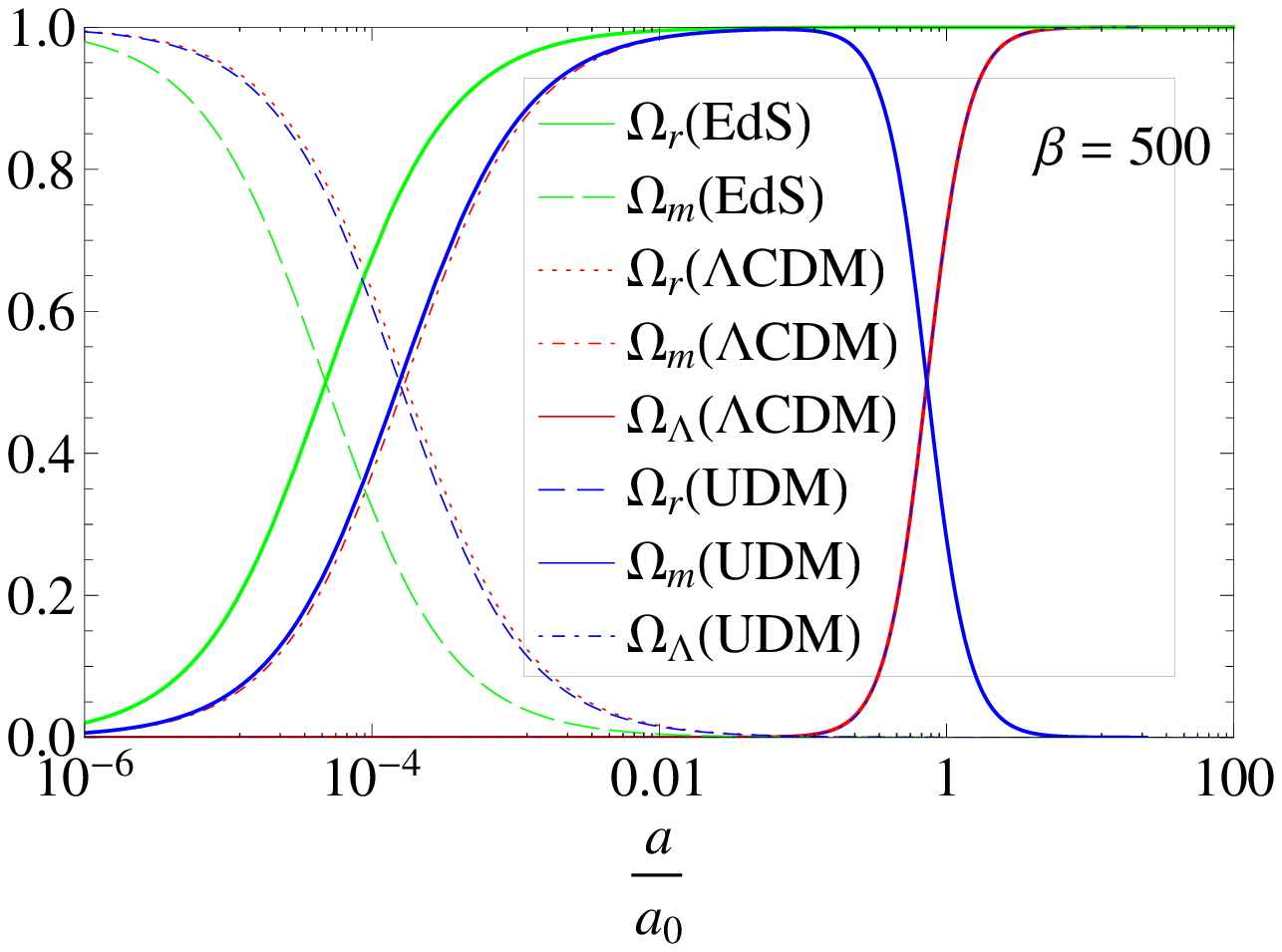}
\caption{Density dependence of the various components for  UDM models with a fast transition ($\beta=500$) occurring at
different times, contrasted with the pure CDM (EdS) and $\Lambda$CDM models, assuming $\Omega_{\Lambda,0}=0.72$ today.
Top panel: $z_{\rm t}=0$; medium panel: $z_{\rm t}=1$; bottom panel: $z_{\rm t}=2$.}
\label{omegadensity}
\end{center}
\end{figure}

Finally, let us consider the evolution of the three components $\rho_{\rm r}$ (radiation) $\rho_{\rm m}$ and
$\rho_\Lambda$,
represented by the dimensionless functions $\Omega_{\rm i}(a)$, Eq.\ (\ref{omegas}). The evolution of these functions
is
shown in Fig.\ \ref{omegadensity} for  UDM models with fast transition ($\beta=500$) occurring at different times,
contrasted with pure the CDM  (EdS) and $\Lambda$CDM models. For UDM and $\Lambda$CDM, it is assumed that today
$\Omega_{\Lambda,0}=0.72$. In a flat Universe, $\Sigma_{\rm i}~\Omega_{\rm i}=1$, and when a component $j$ dominates
$\Omega_{\rm j}\approx 1$, with the other  $\Omega_{\rm i}\approx 0$. It can be seen from these figures that if the
transition is too late (today
in the extreme case $z_{\rm t}=0$) then the matter-radiation equality of the UDM model is basically the same as in a
pure CDM
model, i.e.\ much earlier than in $\Lambda$CDM. In addition, the effective cosmological constant of the UDM  becomes
dominant at a  later time than in $\Lambda$CDM.  Since the matter-transition equality dictates when matter perturbations
 inside the horizon start to grow, while the late dominance of the cosmological constant  slows down this growth, the
matter power spectrum in fast transition UDM models with too late a transition will be at odds with the observed one,
with too much power on small scales.

Conversely, we see from Fig.\ \ref{omegadensity} that if the transition is at $z_{\rm t}\sim 1$ the matter-radiation
equality
is closer to that of $\Lambda$CDM, and it essentially coincides with the latter for an even earlier transition,
$z_{\rm t}\sim 2$. Basically, as long as the transition is at a redshift $z_t$ higher than the one at which
$\Omega_\Lambda=\Omega_{\rm m}$ in $\Lambda$CDM, the late time evolution of  $\Omega_\Lambda$ and $\Omega_{\rm m}$ in
the UDM
models is the same as in $\Lambda$CDM.

\subsection{Angular diameter distance}
\begin{figure}
\begin{center}
\includegraphics[width=1.1\columnwidth]{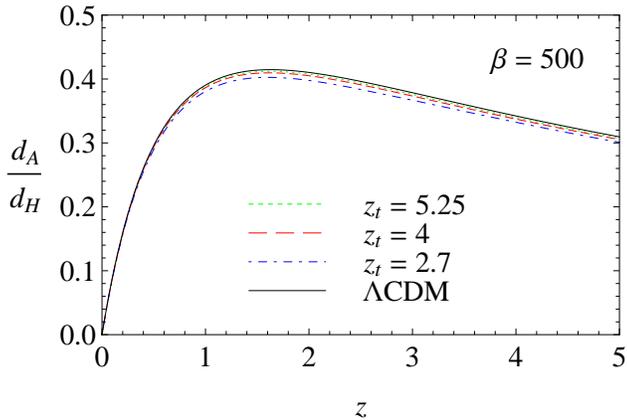}
\caption{The dimensionless angular diameter distance $d_{\rm A}/d_{\rm H}$ for the $\Lambda$CDM and the UDM model with a
fast
transition at $\beta=500$ for different transition redshifts.}
\label{dazvar}
\end{center}
\end{figure}
The angular diameter distance is an important quantity that comes into play in current observations of weak lensing,
Baryon Acoustic Oscillations (BAO) and galaxy clustering and will become even more important for comparing models against the new data that will
become available from surveys such as Dark Energy Survey (DES), Planck and Euclid. It is also relevant to measurements of CMB anisotropies.
In view of this, we now briefly comment on the deviation of the angular diameter distance in our UDM model from that of
$\Lambda$CDM.
\begin{figure}
\begin{center}
\includegraphics[width=1\columnwidth]{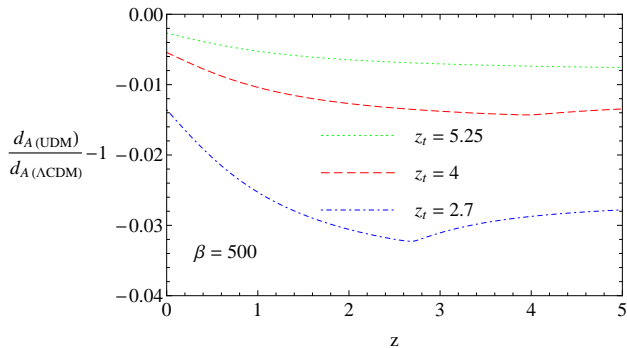}
\caption{Difference between $d_{\rm A}/d_{\rm H}$ for the $\Lambda$CDM and for the UDM model with a fast transition at
$\beta=500$.}
\label{da-1}
\end{center}
\end{figure}
The angular diameter distance, $d_{\rm A}$, is defined as the ratio of the actual size $\triangle x$ of an object and
the
angle $\triangle\theta$ this object subtends orthogonal to the line of sight and can also be expressed as:
\begin{equation}\label{dA}
d_{\rm A}=\frac{1}{1+z}\int^{z}_{0}\frac{dz'}{H(z')}\;,
\end{equation}
 where $H(z)$ is the Hubble function. In Fig.\ \ref{dazvar}, we have show the angular diameter distance normalised with
the Hubble distance, $d_{\rm H}\equiv H^{-1}_{0}$, for the $\Lambda$CDM model and our UDM model with a fast transition
with
$\beta=500$. As it can be seen, the larger the transition redshift, the smaller the departure of $d_{\rm A}$  for the
UDM
with respect to the $\Lambda$CDM is. Notice that $d_{\rm A}$ does not increase indefinitely as $z\rightarrow\infty$; it
turns over at $z\sim 1.5$ and thereafter more distant objects actually appear larger in angular size.

\section{The Jeans scale and the gravitational potential}\label{sec:perts}
\subsection{The Jeans wave number}
We now focus on the Jeans wave number for our UDM model and investigate its behaviour as a function of the speed of
sound, in particular around $\rho=\rho_{\rm t}$, which corresponds to the middle of the transition where the speed of
sound
is at its peak.

\begin{figure}
\begin{center}
\includegraphics[width=0.5\columnwidth]{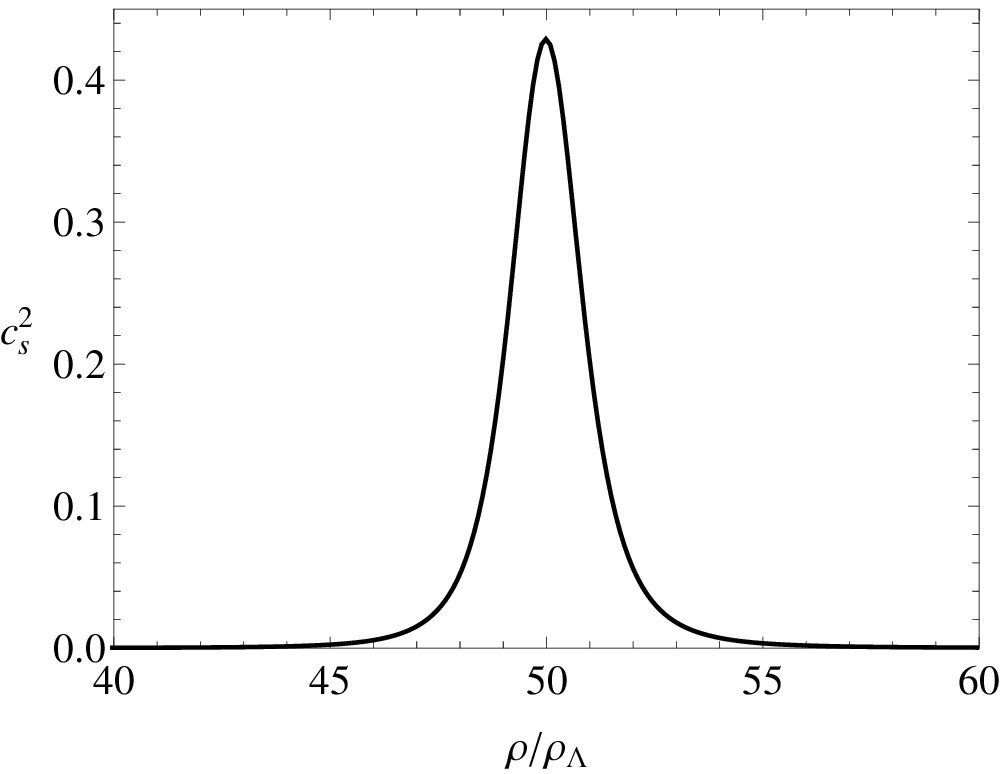}\includegraphics[width=0.5\columnwidth]{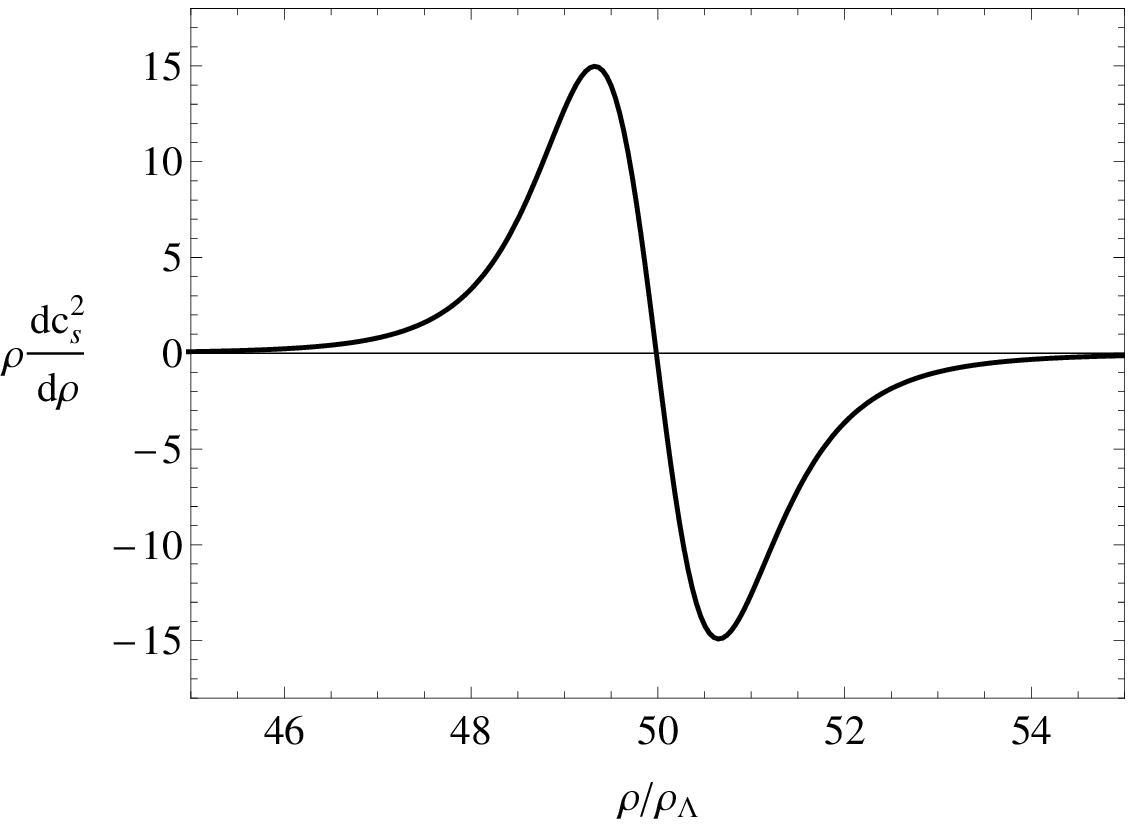}
\caption{Illustrative parametric plots of the speed of sound and $\rho\; dc_{\rm s}^2/d\rho$ as functions of $\rho/\rho_
\Lambda$ with $\rho_{\rm t}/\rho_\Lambda = 50$ ($z_{\rm t}\thicksim4$) and $\beta=500$. The speed of sound reaches its
maximum value at $\rho=\rho_{\rm t}$.}
\label{FigSoS}
\end{center}
\end{figure}

By inspection of Eq.~(\ref{kJ2analytic}) we see that a large $k_{\rm J}^{2}$ can be obtained not only when $c_{\rm s}^2
\to
0$, but when $c_{\rm s}^2$ changes rapidly as well. In other words, when Eq.~(\ref{kJ2analytic}) is dominated by the
$\rho\; dc_{\rm s}^2/d\rho$
term we may say that the EoS is characterised by a fast transition. These two quantities, $c_{\rm s}^2$ and $\rho\;
dc_{\rm s}^2/d\rho$, are depicted in Fig. \ref{FigSoS} as functions of $\rho/\rho_\Lambda$.

Thus, viable adiabatic UDM models can be constructed which do not require $c_{\rm s}^2 \ll 1$  at all times if the
speed of sound goes through a rapid change, a fast transition period during which  $k_{\rm J}^{2}$ can remain large, in
the sense that  $k^{2} \ll k_{\rm J}^{2}$ for all scales of cosmological interest to which the linear perturbation
theory of Eq.~(\ref{equ}) applies.

When we consider a fast transition, it is interesting to compare the term $\rho\,{d}c_{\rm s}^2/{d}\rho$ with the
remaining ones contained in the squared brackets of Eq.~(\ref{kJ2analytic}) for the Jeans wave number, that is:
\begin{equation}
 \mathcal{B}= \frac{1}{2}(c_{\rm s}^2 - w) + \frac{3(c_{\rm s}^2 - w)^2 - 2(c_{\rm s}^2 - w)}{6(1 + w)} + \frac{1}{3}\;.
\end{equation}

\begin{figure}
\begin{center}
\includegraphics[width=0.48\columnwidth]{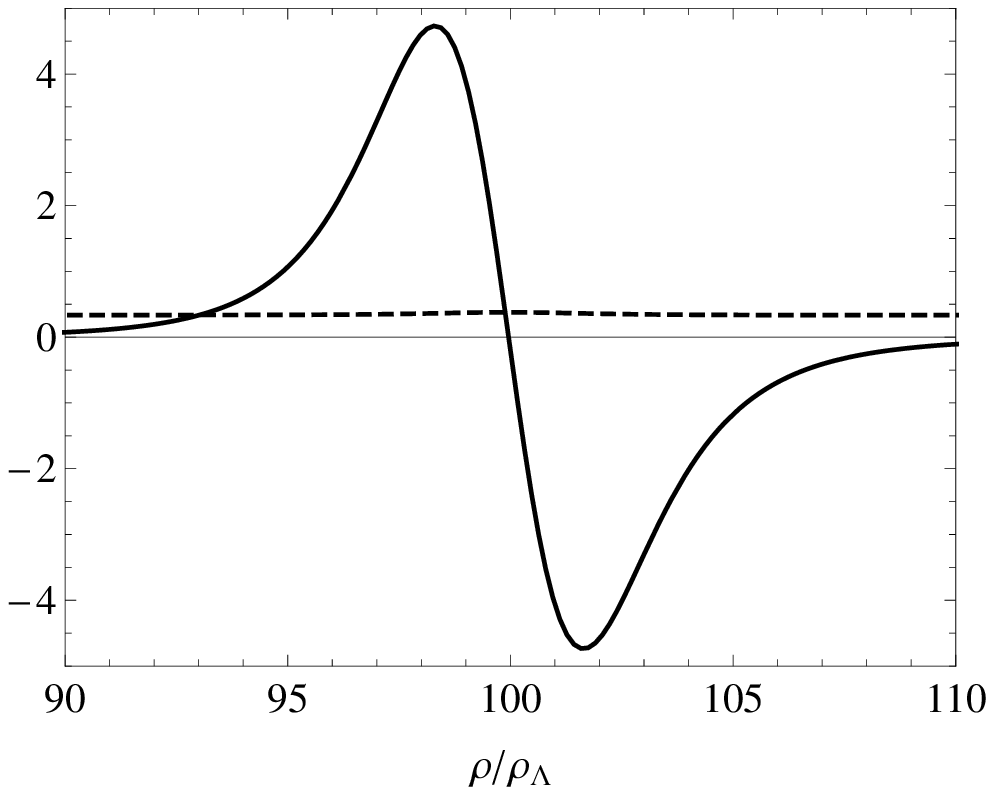}
\includegraphics[width=0.497\columnwidth]{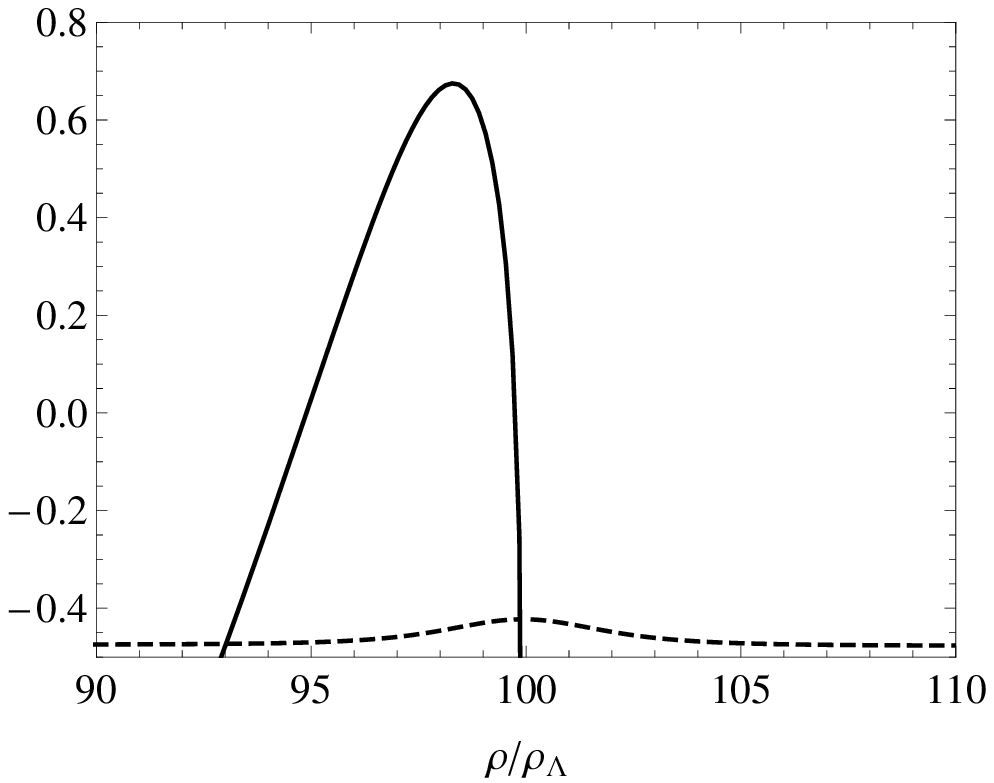}
\includegraphics[width=0.5\columnwidth]{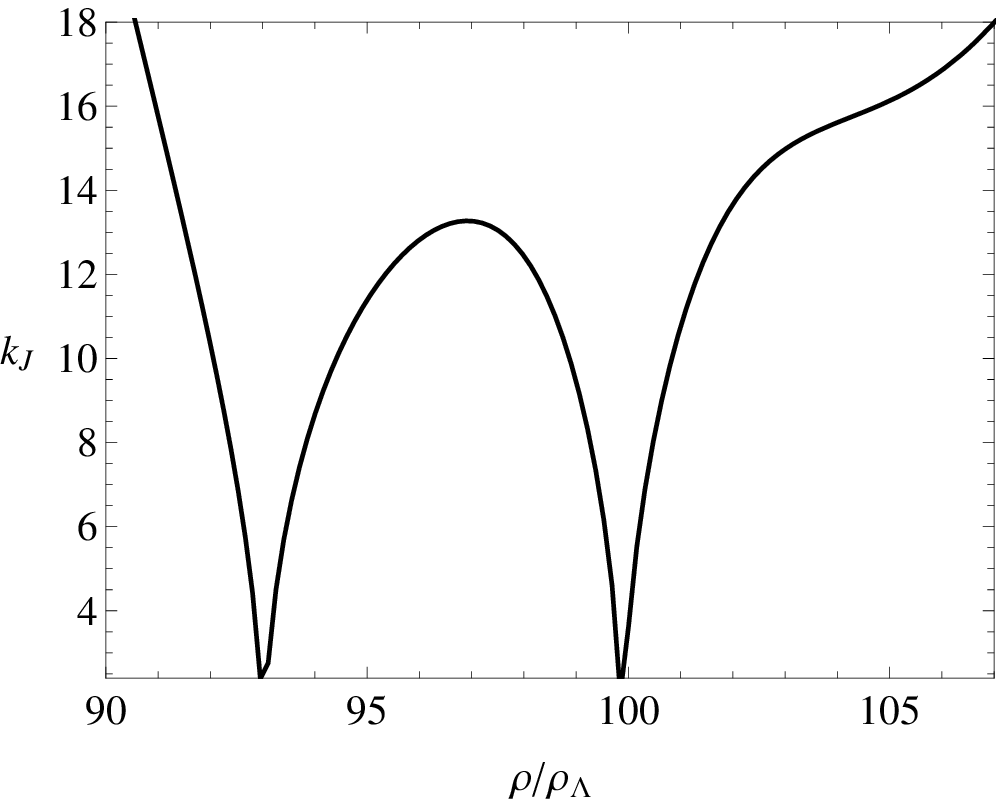}
\caption{Left-top panel: evolution of $\rho\; dc_{\rm s}^2/d\rho$ (solid line) and $\mathcal{B}$ (dashed line) as
functions of $\rho/\rho_\Lambda$. Right-top panel: evolution of $\rho\; dc_{\rm s}^2/d\rho$ (solid line) and $\mathcal{
B}$ (dashed line) as functions of $\rho/\rho_\Lambda$ in a logarithmic scale. Bottom panel: evolution of the Jeans wave
number $k_{\rm J}$ as a function of $\rho/\rho_\Lambda$ where $k_{\rm J}$ is in $h$ Mpc$^{-1}$ units. The choice of
parameters
is  $\rho_{\rm t}/\rho_\Lambda = 100$ ($z_{\rm t}\thicksim5.34$) and $\beta=500$.}
\label{FigSoS2}
\end{center}
\end{figure}
In Fig.\ \ref{FigSoS2} we plot $\rho\,{d} c_{\rm s}^2/{ d}\rho$, $\mathcal{B}$ and the Jeans wave number $k_{\rm J}$ as
functions of $\rho/\rho_\Lambda$.

\begin{figure}
\begin{center}
\includegraphics[width=0.49\columnwidth]{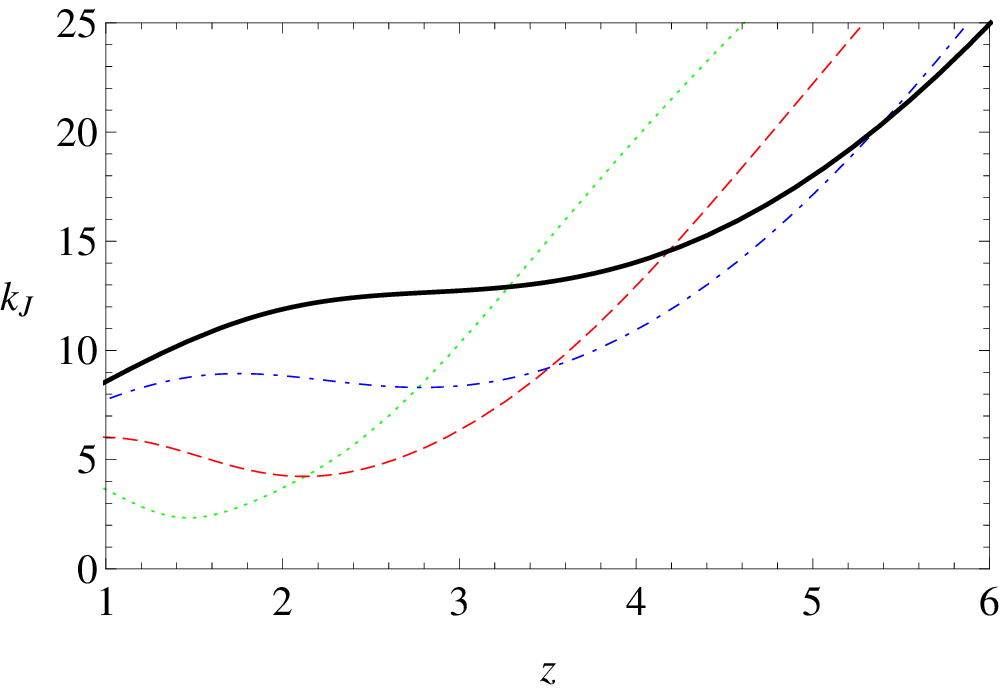}
\includegraphics[width=0.49\columnwidth]{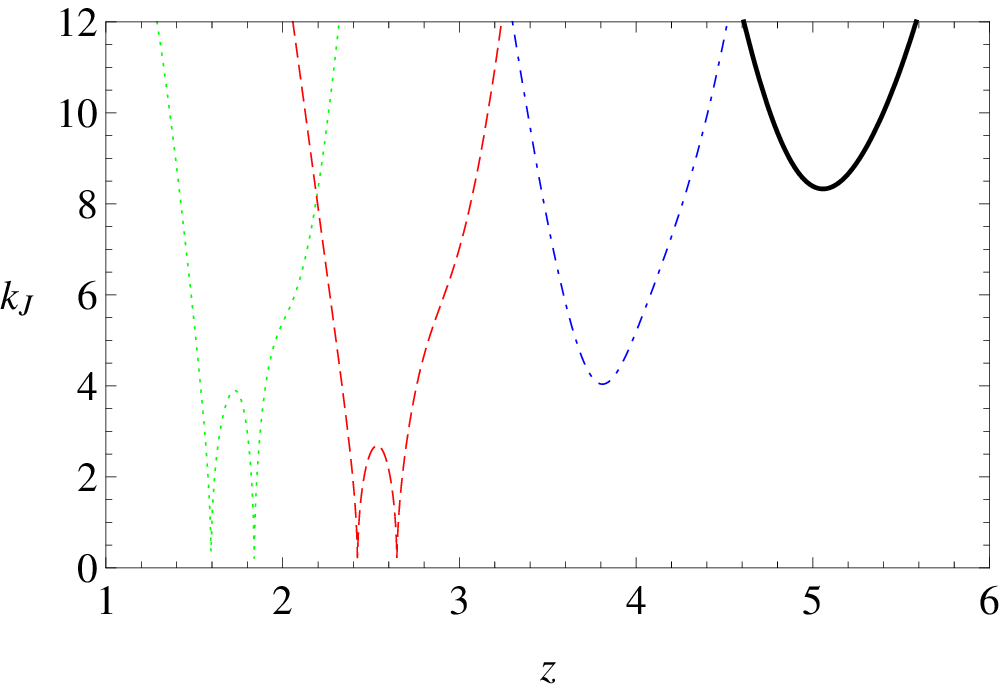}
\includegraphics[width=0.49\columnwidth]{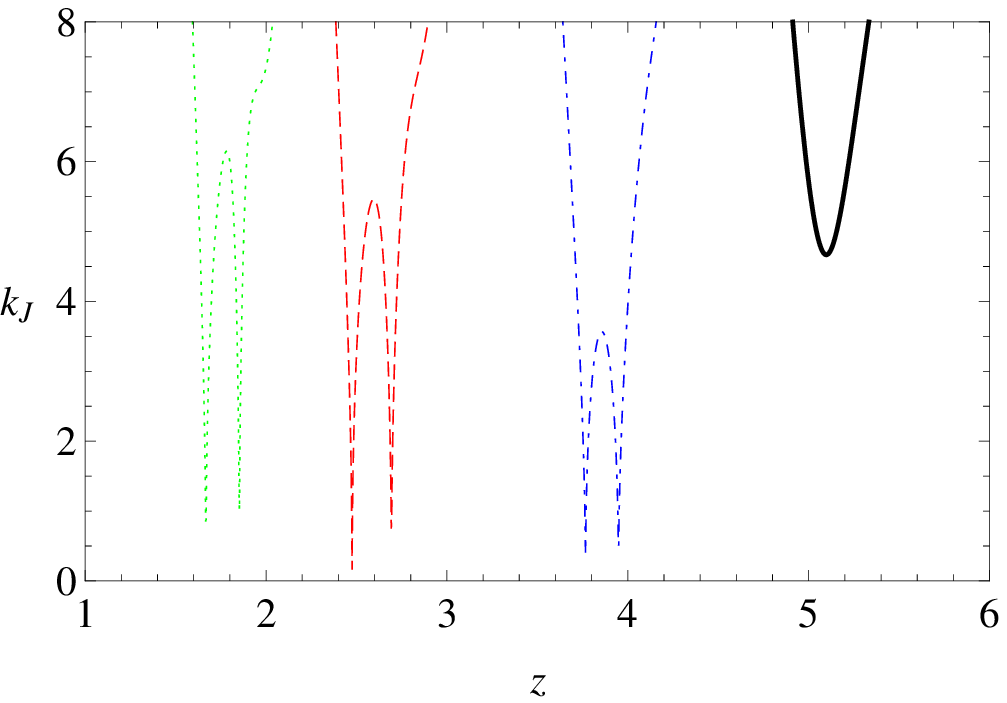}
\includegraphics[width=0.49\columnwidth]{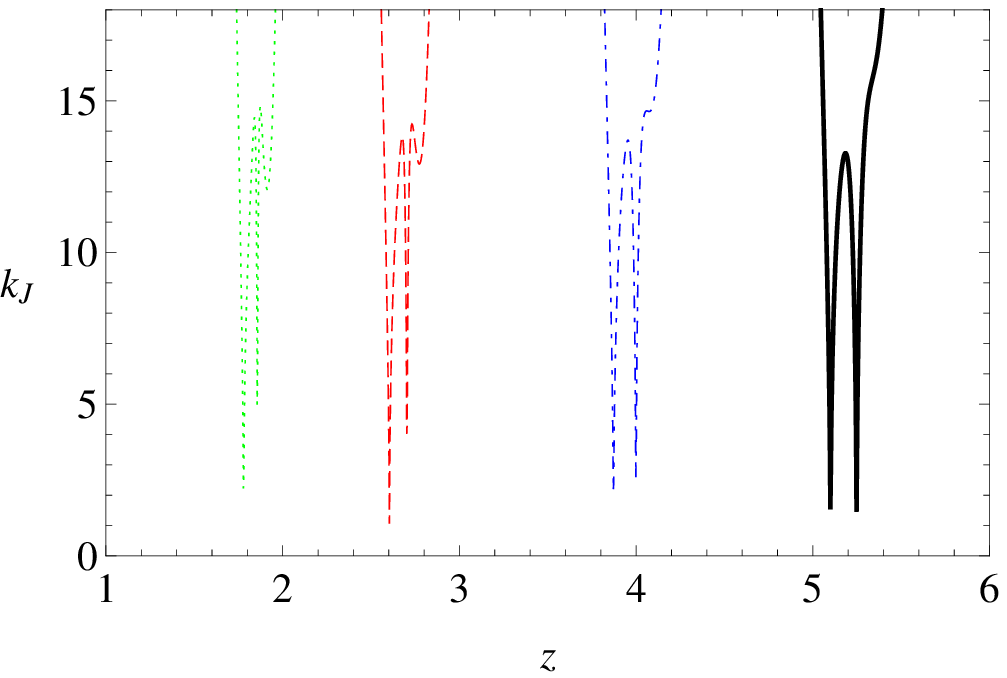}
\caption{The Jeans wave number $k_{\rm J}$ (in $h$ Mpc$^{-1}$ units) for $\beta=10$ (left-top panel), $\beta = 50$
(right-
top panel), $\beta = 100$ (left-bottom panel) and $\beta = 500$ (right-bottom panel). In each panel, the three dotted,
dash-dotted and dashed lines correspond to $\rho_{\rm t}/\rho_\Lambda = 10, 20, 50$, respectively; the black solid line
corresponds to $\rho_{\rm t}/\rho_\Lambda = 100$.}
\label{Figkjvsz}
\end{center}
\end{figure}
From Fig.\ \ref{FigSoS2} we learn that $\rho\, {d} c_{\rm s}^2/{d}\rho$ is negative for $\rho > \rho_{\rm t}$, then for
$\rho\approx \rho_{\rm t}$ it increases becoming positive and intersecting the $\mathcal{B}$ curve a first time for
$\rho < \rho_{\rm t}$. For smaller values of the energy density, $\rho\, {d} c_{\rm s}^2/{d}\rho$ decreases again to
zero, again intersecting the $\mathcal{B}$ curve. Since the difference
between the two curves is very large, we have depicted them in the
right top panel of Fig.\ \ref{FigSoS2} choosing a logarithmic
scale. For this reason, the negative part of $\rho\, { d} c_{\rm s}^2/{ d}\rho$ has been omitted. It can be also seen
that the relative maximum of the Jeans wave number between the two zeros of the curve approximately corresponds to the
point where ${d}c_{\rm s}^2/{d}\rho$ reaches its maximum value.

The place where the curves $\rho\, { d} c_{\rm s}^2/{d}\rho$ and
$\mathcal{B}$ intersect correspond to the vanishing points of the
Jeans wave number $k_{\rm J}$, as it can be seen in the bottom
panel of Fig.\ \ref{FigSoS2}. In general, around these points the
corresponding Jeans length becomes very large, possibly causing
all sort of problems to perturbations, with effects on structure
formation in the UDM model.
Defining $c_{\rm s}^2$ as in Eq.~(\ref{varie2}), in
Fig.\ \ref{Figkjvsz} we plot the Jeans wave number as a function of
the redshift for various values of $\beta$ and
$\rho_{\rm t}/\rho_{\Lambda}$. In this figure, for sufficiently high
$\beta$ we note that {\it i)} in general the Jeans wave number
becomes larger, with a vanishingly small Jeans length before and after the transition, and {\it ii)} it becomes
vanishingly small for
extremely short times, so that the effects caused by its vanishing
are sufficiently negligible, as we are going to show in the next
subsection when we analyse the gravitational potential $\Phi$.

To conclude, we shall make some comments on building phenomenological UDM (or DE) fluid models intended to represent
the homogeneous FLRW background and its linear perturbations. A fast transition in a fluid model could be characterised
by a value of $c_{\rm s}^{2}>1$ during the transition. Notice that it is standard to refer to the parameter $c_{s}^{2}$ as the speed of sound because this is what it would be if Eq.\ (7) was a simple wave equation. In reality, $c_{s}^{2}$ is not the speed of signal propagation because in Eq.\ (7) we also have a potential term $\theta''/\theta$ and, if there is any signal propagation, this would only happen on scales smaller than the Jeans length, and with a speed given by the group velocity \citep{Brillouin}. Therefore, having $c_{s}^{2}>1$ does not raise \emph{per se} any issue with respect to causality, see \cite{Brillouin, Babichev:2007dw}. More specifically on our model,  Eq.~(\ref{equ}) is the Fourier component of a wave equation with potential
${\theta^{\prime\prime}}/{\theta}$, and the latter does not allow propagation for $k \ll k_{\rm J}$. Therefore, we note in addition that we can always
build our fluid model in such a way that all scales smaller than the Jeans length $\lambda \ll \lambda_{\rm J}$
 correspond to those in the non-linear regime, for which this model may  not apply. In order to study the behaviour of
the perturbations of a UDM model at these scales, we would have to go beyond the perturbative regime investigated here.
That would possibly imply to build a more refined fluid model that could maintain causality and at the same time be
able to deal with the increased complexity of small scale non-linear physics.

\subsection{The gravitational potential}

The differential equation that governs the behaviour of the gravitational potential $\Phi$ in our model in terms of the
scale factor $a$ is obtained from Eq.~(\ref{equ}):

\begin{multline}\label{diff-eq_Phi}
\frac{d^2 \Phi(\textbf{k},a)}{d a^2}+\left(\frac{1}{\mathcal{H}} \frac{d \mathcal{H}}{d a} + \frac{4}{a}+ 3
\frac{c_{\rm s}^2}
{a}\right)\frac{d \Phi(\textbf{k},a)}{d a}+\\
\left[ \frac{2}{a \mathcal{H}}\frac{d \mathcal{H}}{d a} + \frac{1}{a^2}(1+3 c_{\rm s}^2)+\frac{c_{\rm s}^2 k^2}{a^2
\mathcal{H}^2}
\right] \Phi (\textbf{k},a) =0\;,
\end{multline}
where $\mathcal{H}= \frac{da}{d\eta}/a$ is the conformal time Hubble function. Also, $\mathcal{H}=aH$, $c_{\rm s}^2$ is
given in Eq.~(\ref{varie2}) and we have assumed plane-wave perturbations
$\Phi({\bf x},a) \propto  \Phi(\textbf{k},a) \exp\left(i{\bf k} \cdot {\bf x}\right)$ of comoving wave-number
$k\equiv |\bf{k}|$.

On the other hand, the $\Lambda$CDM gravitational potential, $\Phi_{\Lambda}$, solves Eq.~(\ref{diff-eq_Phi}) for
$c_{\rm s}=0$:

\begin{equation}\label{PhiLa}
\frac{d^2 \Phi_{\Lambda}}{d a^2}+\left(\frac{1}{\mathcal{H}} \frac{d \mathcal{H}}{d a} + \frac{4}{a}\right)
\frac{d\Phi_{\Lambda}}{d a}+\left[ \frac{2}{a \mathcal{H}}\frac{d \mathcal{H}}{d a} +  \frac{1}{a^2}\right]
\Phi_{\Lambda} =0\;,
\end{equation}
where
\begin{equation}\label{HLCDM}
\mathcal{H}^2 = H_0^2\left(\Omega_{\Lambda,0}a^2 + \Omega_{m,0}a^{-1}\right)\;,
\end{equation}
if we assume that the energy density of radiation is negligible and ignore the contribution of the baryon matter.

Comparing Eqs.~(\ref{diff-eq_Phi}) and (\ref{PhiLa}) we see that the gravitational potential in our UDM model has the
same evolution as in a $\Lambda$CDM universe in the limit when $c_{\rm s} \to 0$. In particular, both models behave as
an EdS model at early times, so that for UDM $c_{\rm s} \to 0$ and in both models $\mathcal{H}^2\sim a^{-1}$, and $\Phi$
is constant in this regime, as in an EdS universe, as is well known.

The normalised initial conditions are
$\Phi_{\Lambda}(\textbf{k};a_{\rm rec}) = 1$ and $\left.
d\Phi_\Lambda/da\right|_{a_{\rm rec}} = 0$, where $a_{\rm rec}$
stands for the scale factor at recombination time. Since the class
of UDM models we consider here is constructed to behave as the
$\Lambda$CDM model in the early Universe, we thus set the same
initial conditions for both the UDM and the $\Lambda$CDM
gravitational potentials.

\begin{figure}
\begin{center}
\includegraphics[width=0.4948\columnwidth]{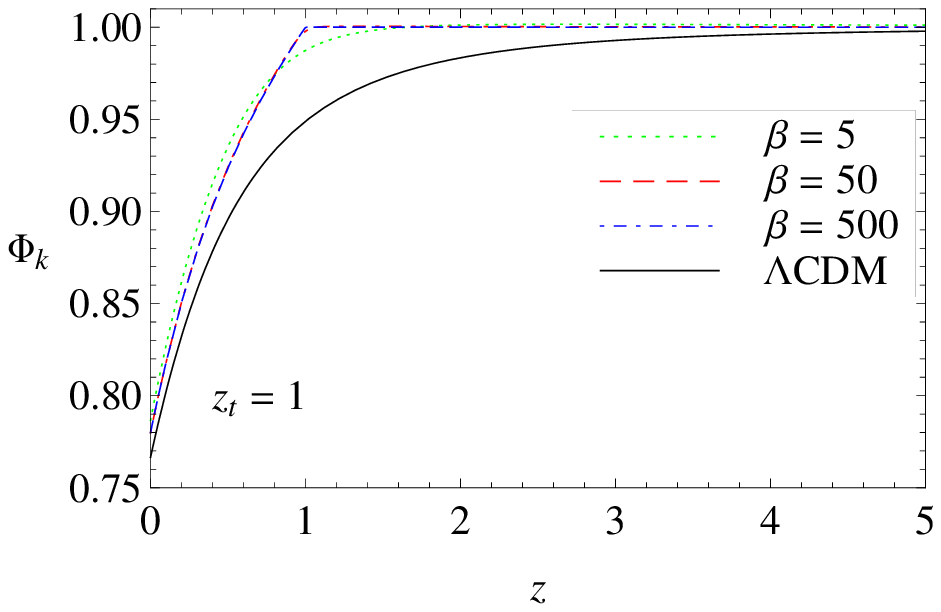}
\includegraphics[width=0.4948\columnwidth]{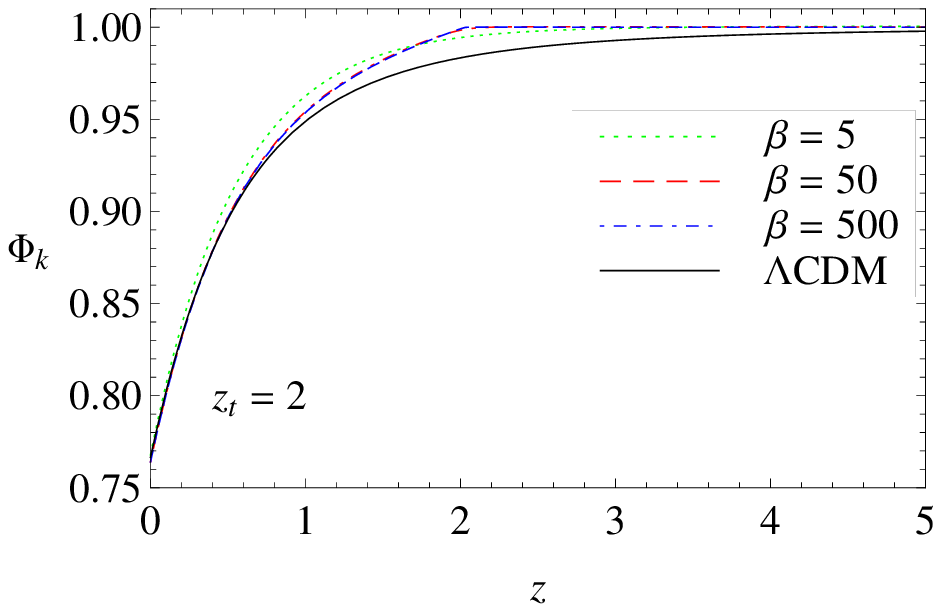}
\includegraphics[width=0.4948\columnwidth]{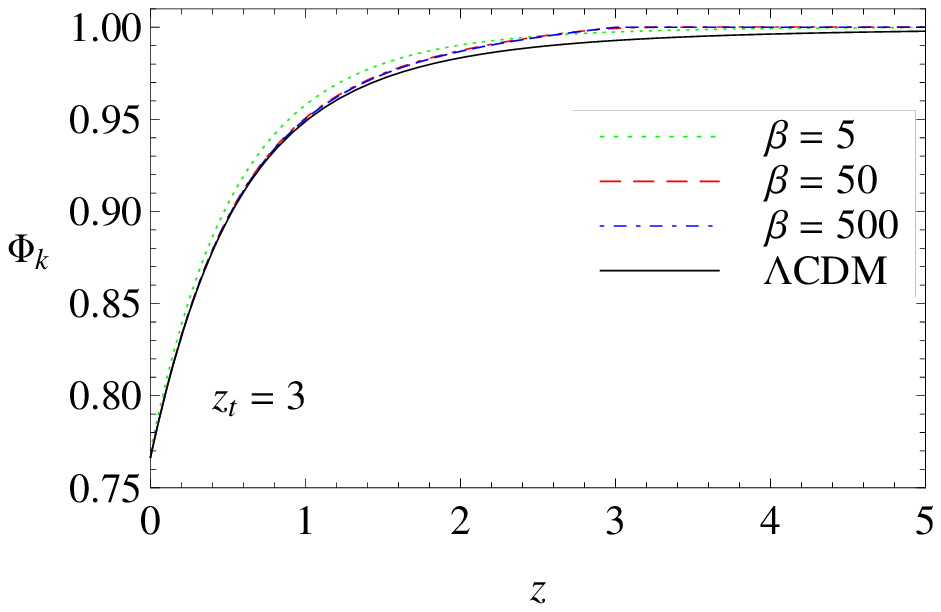}
\includegraphics[width=0.4948\columnwidth]{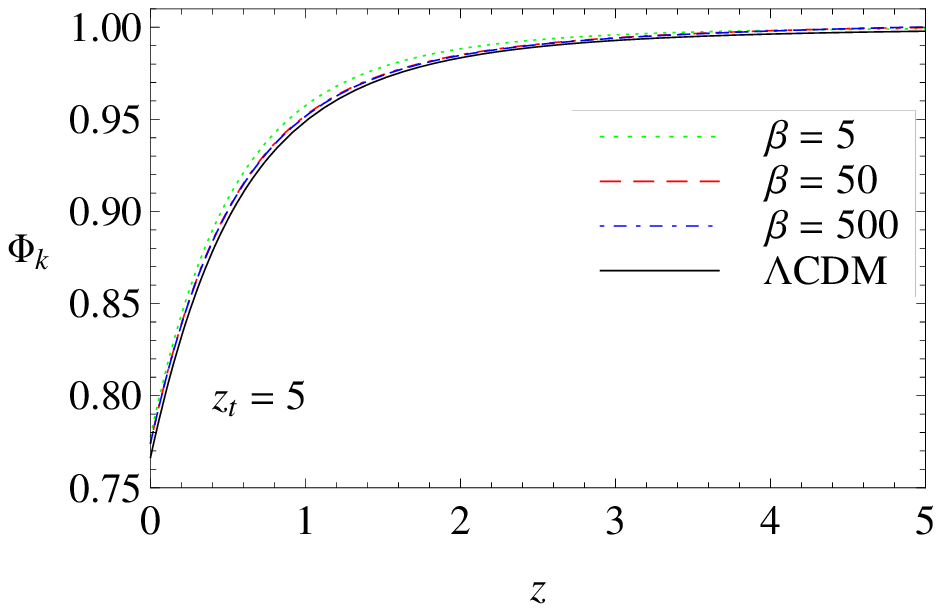}
\caption{Illustrative plots of the gravitational potential $\Phi(\textbf{k};z)$ as a function of the redshift $z$ for
$\Lambda$CDM and for our UDM model for $k=0.2$ $h$ Mpc$^{-1}$ and different values of $\beta$ and $z_{\rm t}$. The
black solid line corresponds to the gravitational potential in the $\Lambda$CDM model with $\Omega_{\Lambda,0} = 0.72$.}
\label{gpvsz}
\end{center}
\end{figure}

In $\Lambda$CDM, the background evolution causes a gradual time evolution of the gravitational
potential when the cosmological constant starts to dominate \citep{Hu:1995em}; this causes an ISW effect. On the other
hand, in our UDM model the evolution of the gravitational potential is determined by the background and the perturbative
 evolution of the single dark fluid and, crucially, by the adiabatic speed of sound $c_{\rm s}^{2}$. The gravitational
potential stays  constant before the transition, during which a sudden rapid evolution of $\Phi$ is induced. The
subsequent evolution is in general scale dependent: for scales $k>k_{\rm J}$, $\Phi$  oscillates and decays; for larger
scales, $k<k_{\rm J}$, the evolution of $\Phi$ becomes scale independent and is governed by the evolution of the
background,
 mainly by $H$ and to a small extent by $c_{\rm s}^{2}$, and $\Phi$ approaches its $\Lambda$CDM behaviour in a way that
depends mainly on the rapidity $\beta$ and from the epoch of the transition, $z_{\rm t}$. In general, we expect an ISW
effect
starting from the transition which can be very different from the $\Lambda$CDM one, cf.\ \cite{Piattella:2009kt}. We show
the behaviour of $\Phi$ in Fig.\ \ref{gpvsz}, where we explore its dependence on the background parameters $\beta$
and $z_{\rm t}$ (or, equivalently, $a_{\rm t}$).

We already know from the evolution of $w$ in Fig.\  \ref{wvszFig2} that for $\beta < 200$ the transition is not that
fast. This is also apparent in Fig.\ \ref{gpvsz}, where we have plotted the
normalised gravitational potential
$\Phi_{\rm k}(z)=\Phi(\textbf{k};z)/\Phi(\textbf{0};10^{3})$ as a
function of the redshift $z$ for $k=0.2$ $h$ Mpc$^{-1}$ and
different values of $\beta$ and $z_{\rm t}$. For wavenumbers
$k>k_{\rm nl}\simeq0.2$ $h$ Mpc$^{-1}$, we expect non-linear matter
over-densities contributions to
the evolution of the gravitational potential to become important.

As expected, for large enough values of $\beta$ the gravitational potential is practically constant in time,
corresponding to a pure matter (EdS) background evolution, until $z\sim z_{\rm t}$. From $z < z_{\rm t}$ onwards, its
value decreases very fast until it finally approaches the gravitational potential of the $\Lambda$CDM model. The
expected strong ISW effect caused by this behaviour would be mostly due to the particular expansion history and can be
very different from the $\Lambda$CDM one. As a matter of fact, these differences are smaller for $z_{\rm t} > 2$ and
become smaller and smaller at higher transition redshifts, until for $z_{\rm t} \sim 100$ and larger the fast
transition UDM models practically become indistinguishable from $\Lambda$CDM cf.\ \cite{Piattella:2009kt}.

As we have already pointed out, our UDM model allows the value
$w=-1$ for $a\rightarrow\infty$, i.e, it admits an effective
cosmological constant energy density $\rho_{\Lambda}$ at late
times. Hence, if we wanted to compare the predictions of our UDM model
with observational data, we would follow the prescription given in
\citep{Piattella:2009kt}, where the density contrast is $\delta\equiv \delta\rho/\rho_{\rm A}$ and
$\rho_{\rm A}=\rho-\rho_{\Lambda}$ is the clustering "aether" part of the
UDM component \citep{Ananda:2005xp,Linder:2008ya}. In UDM models gravity is described by General Relativity, but to
link the density contrast with the gravitational potential at scales much smaller than the cosmological horizon we only
need the Newtonian Poisson equation. For $z < z_{\rm rec}$, where
$z_{\rm rec}$ is the recombination redshift ($z_{\rm rec} \approx 10^{3}$) we then have:
\begin{equation}\label{delta}
 \delta\left(\textbf{k};z\right)  = -\frac{2 k^{2}\Phi(\textbf{k};z)\left(1 + z\right)^2}{\rho_{\rm A}}\;.
\end{equation}

To conclude, we have argued that for an early enough fast transition with $\beta>500$ and $z_{\rm t} > 2$ our UDM model
should be compatible with observations. On the other hand, a study of the matter and CMB power spectra is needed to
study the viability of models with $10 \lesssim  \beta<500$, and those with $\beta>500$ and $z_{\rm t} < 2$. We shall
undertake this work in the future.

\section{Conclusions}\label{sec:concl}

 UDM models, when compared with the standard DM + DE scenario, specifically $\Lambda$CDM, are in principle interesting
because  the dynamics of the Universe can be described with a single component in the matter sector which triggers the
accelerated expansion at late times and is also able to cluster and produce a satisfactory structure formation. The
challenge for UDM models is however to satisfy observational constraints while maintaining features that can make then
distinguishable from $\Lambda$CDM, otherwise they lose interest \citep{Sandvik:2002jz}. In this paper we have
introduced and  examined a new  class of UDM models with a fast transition between an early matter era and a late
$\Lambda$CDM-like phase, building on previous work \citep{Piattella:2009kt,Bertacca:2010mt}.

First, in Sec.~\ref{sec:udmmodels}, we have introduced some generalities of UDM models. In Sec.~\ref{sec:3p} we have
considered
three possible
prescriptions for building phenomenological UDM models, with the aim of obtaining  models in which all the variables of
interest can be expressed analytically, so that in principle they could be  implemented into numerical codes such as
CAMB \citep{Lewis:1999bs} and CLASS \citep{Lesgourgues:2011re} while maintaining the code efficiency. Indeed, in
comparing models with observational data this is of crucial importance in view of likelihood analysis, and a major
motivation for the UDM model presented here:  modifying these numerical codes to deal with fast transition UDM models
while maintaining their efficiency is in general a non trivial task \citep{Piattella:2009kt},  thus having as many
variables as possible expressed analytically simplifies the task considerably.
 While in \cite{Piattella:2009kt} the fast transition was introduced in the equation of state, we have shown in
Sec.~\ref{sec:3p}
that the best prescription to proceed as much as possible analytically is to assume a specific evolution of the energy
density of UDM.

A general feature of UDM models is in the possible difference of the expansion history with that of $\Lambda$CDM,
causing, among other features, a strong ISW effect incompatible with observations.
In addition, in UDM models
the  effective speed of sound  may become significantly different from zero. This corresponds, in general, to the
appearance of a Jeans length (or sound horizon) below which the dark fluid cannot cluster and which, if large enough,
can cause a strong evolution in time of the gravitational potential, preventing structure formation at small scales.
In building satisfactory UDM parametric models it is therefore crucial to find the region in parameter space where the
Jeans length remains small enough, well beyond the linear regime that we explore here.

UDM models with a fast transition, first introduced in \cite{Piattella:2009kt,Bertacca:2010mt} are a viable and
interesting alternative to $\Lambda$CDM because they seem to survive observational tests while maintaining interesting
features.

The new general phenomenological UDM models we have introduced in Sec.~\ref{sec:tghmodel} (following the prescription
obtained in
Sec.~\ref{sec:3p}) are characterised by  a  fast transition between a standard matter era  and a  post-transition
epoch
described  by an ``affine model"  \citep{Ananda:2005xp,Ananda:2006gf,Balbi:2007mz,Quercellini:2007ht,Pietrobon:2008js}
with affine parameter $\alpha$. We have then focused on the  $\alpha=0$ case, which represents a sudden transition to a
$\Lambda$CDM-like late evolution. In constructing these models in practice,  we have to choose a step-like function
representing the fast transition. In doing this, for physical reasons we want to maintain the condition
$c_{\rm s}^{2}>0$
at all times: after  carrying out an extensive study over many possibilities \citep{Bracewell} we have chosen the
function in  Eq.\ (\ref{Hc}) as the only one we found
 that complied with this condition. In  Sec.\ 4 we have also compared the angular diameter distance between $\Lambda$CDM
and our UDM with fast transition, finding small differences of the order of percent when the transition is
fast enough.

Finally, in Sect.~\ref{sec:perts}, in order to study the viability of our UDM model, we have carried out  a  study of
the functional
form of the Jeans scale in adiabatic UDM perturbations. In doing so, we have found analytical expressions for the
quantities involved in the Jeans wave number and have shown that our model presents a small Jeans length even when a
non-negligible sound speed is present. Subsequently, we have analysed the properties of perturbations in our model,
focusing on the evolution of the effective speed of sound, the Jeans scale and the gravitational potential. In general,
in building a phenomenological model, we have chosen its parameter values  in order to always satisfy the condition
$k \ll k_{\rm J}$ for all $k$ of cosmological interests to which linear theory applies. In this way, we have been able
to set theoretical constraints on the parameters of the model, predicting sufficient conditions for the model to be
viable.
 We have argued that for large enough values of the rapidity $\beta$ of the transition and $z_{\rm t}$ our model should
be
compatible with observations. Overall, we have found results for our new UDM model similar to those in
\cite{Piattella:2009kt} but, given that we have started by prescribing a specific energy density evolution for the UDM
component rather than from a fast transition in the equation of state, this is a non trivial outcome.

Computing the CMB and the matter power spectra for our model, for a wide range of parameters values, as well as a full
likelihood analysis for this model and its parameters, including $\alpha\neq0$, will be the subject of a forthcoming
work. Other possible extensions of the work presented here could aim at including isocurvature (entropy) perturbations, following the prescription of \cite{Pietrobon:2008js} for the ``affine model", an ``affine" post-transition era (a possibility that we have considered in Sec.\ 4.1), as well as formulating our model in terms of a non-standard scalar field, along the lines of \cite{Bertacca:2010mt}.

\section*{Acknowledgements}
The authors  thank Robert Crittenden, Marc Manera and Francesco Pace for useful discussions. MB is
supported by the STFC (grant no. ST/H002774/1), RL by the
Spanish Ministry of Economy and Competitiveness
through research projects FIS2010-15492 and Consolider
EPI CSD2010-00064, the University of the Basque Country UPV/EHU under
program UFI 11/55 and also by the ETORKOSMO special research action. ARF is supported by the `Fundaci\'on Ram\'on
Areces'.

\bibliography{BFastUDM}
\bibliographystyle{mn2e}

\label{lastpage}

\end{document}